\documentstyle{l-aa}

\input{psfig.tex}

\newcommand{\Edot}{$\lg\dot{E}$}

\newcommand{\Pdot}{$\dot{P}\times 10^{-15}$}

\newcommand{\ApJ}{ApJ}
\newcommand{\ApJS}{ApJS}

\newcommand{\Nat}{Nat}
\newcommand{\Aa}{A\&A}
\newcommand{\AaS}{A\&AS}
\newcommand{\MNRAS}{MNRAS}

\begin{document}

   \thesaurus{16                        
               (16.14.1,                
                                        %
                                       %
               16.13.1;                
               19.50.1;                
               24.03.1;                
               19.16.1)                
              }

\title{The X-ray luminosity of rotation-powered neutron stars}

\author{W.~Becker \and J.~Tr\"umper}

\offprints{web@mpe-garching.mpg.de}

\institute{Max-Planck-Institut f\"ur Extraterrestrische Physik,
            85740 Garching bei M\"unchen, Germany }
 
\date{Received: 2.~January 1997 / Accepted 30 April 1997} 
\maketitle
 
\begin{abstract}
 As a result of recent observations with ROSAT and ASCA the number of 
 rotation-powered pulsars seen at X-ray energies has increased substantially. 
 In this paper we review the phenomenology of the observed X-ray emission 
 properties. At present 27 pulsars are detected, representing a wide range of
 ages ($10^3 - 7\times 10^9$ yrs), magnetic field strength ($10^8-10^{13}$ G) 
 and spin periods ($1.6 - 530$ ms). Despite these dispersions in parameters all 
 pulsars show an X-ray luminosity closely correlated with the rotational energy 
 loss. This suggests that most of the observed X-rays are produced by magnetospheric 
 emission originating from the co-rotating magnetosphere. Only for three middle 
 aged pulsars (PSR 0656+14, Geminga and PSR 1055-52) and probably for the 
 Vela-pulsar an additional thermal component is detected which can be 
 attributed to thermal emission from the neutron stellar surface.

 \keywords{Pulsars: individual (PSR 0531+21, 0833-45, 0633+17, 1706-44
  1509-58, 1951+32, 1046-58, 1259-63, 1823-13, 1800-21, 1929+10, J0437-4715,
  1821-24, J2124-3358, 0656+14, 0540-69, 1957+20, 0950+08, 1610-50, 0538+28,
  J1012+5307, 1055-52, 0355+54, 2334+61, J0218+4232, 0823+16, J0751+1807) -- 
  X-rays: general -- Stars: neutron -- Stars: binaries: general}
\end{abstract}

\section{Introduction}
 The nearly 750 radio pulsars detected so far are interpreted as rapidly 
 spinning and strongly magnetized neutron stars which are radiating at the 
 expense of rotational energy (Pacini 1967). 
 Although the popular model of magnetic braking provides plausible estimates 
 for the neutron star magnetic dipole component $B_\perp$, its braking energy
 $\dot{E}$ and characteristic age $\tau$, it does not provide any detailed 
 information about the physical mechanism which operates in the pulsar 
 magnetosphere and which is responsible for the broad band spectrum from the 
 radio to the X- and gamma-ray bands. As a consequence, there exist a number 
 of magnetospheric emission models, but no accepted theory. A recent review 
 of the observational and theoretical situation is given in the proceedings 
 of the IAU Colloquium 160 (Johnston, Walker \& Bailes 1996). 
 
 In this paper we review the observed emission properties of rotation-powered 
 neutron stars in the soft X-ray band. This radiation has been attributed to
 various thermal and non-thermal emission processes including

 \begin{itemize}

  \item  Non-thermal emission from relativistic particles accelerated in 
         the pulsar magnetosphere. The emission is characterized by a 
         power-law spectrum (cf.~Michel 1991 and references therein).

  \item  Photospheric emission from the hot surface of a cooling neutron 
         star. In this case a modified black-body spectrum and smooth, low amplitude 
         variations with rotational phase are expected (cf.~Greenstein \& Hartke 
         1983; Romani 1987; Pavlov et al.~1995).

  \item  Thermal emission from the neutron star's polar caps which are 
         heated by the bombardment of relativistic particles streaming back 
         to the surface from the pulsar magnetosphere (Kundt \& Schaaf 1993; 
         Zavlin, Shibanov \& Pavlov 1995; Gil \& Krawczyk 1996).
  
  \item Extended emission from a pulsar driven synchrotron nebula 
         (cf.~Michel 1991 and references therein).

  \item  Soft X-ray emission from a relativistic pulsar wind or from a 
         possible interaction of that wind with the interstellar medium or with 
         a close companion star (bow shock nebula) (cf.~Arons \& Tavani 1993).
\end{itemize}

 \begin{figure*}
  \centerline{\psfig{figure=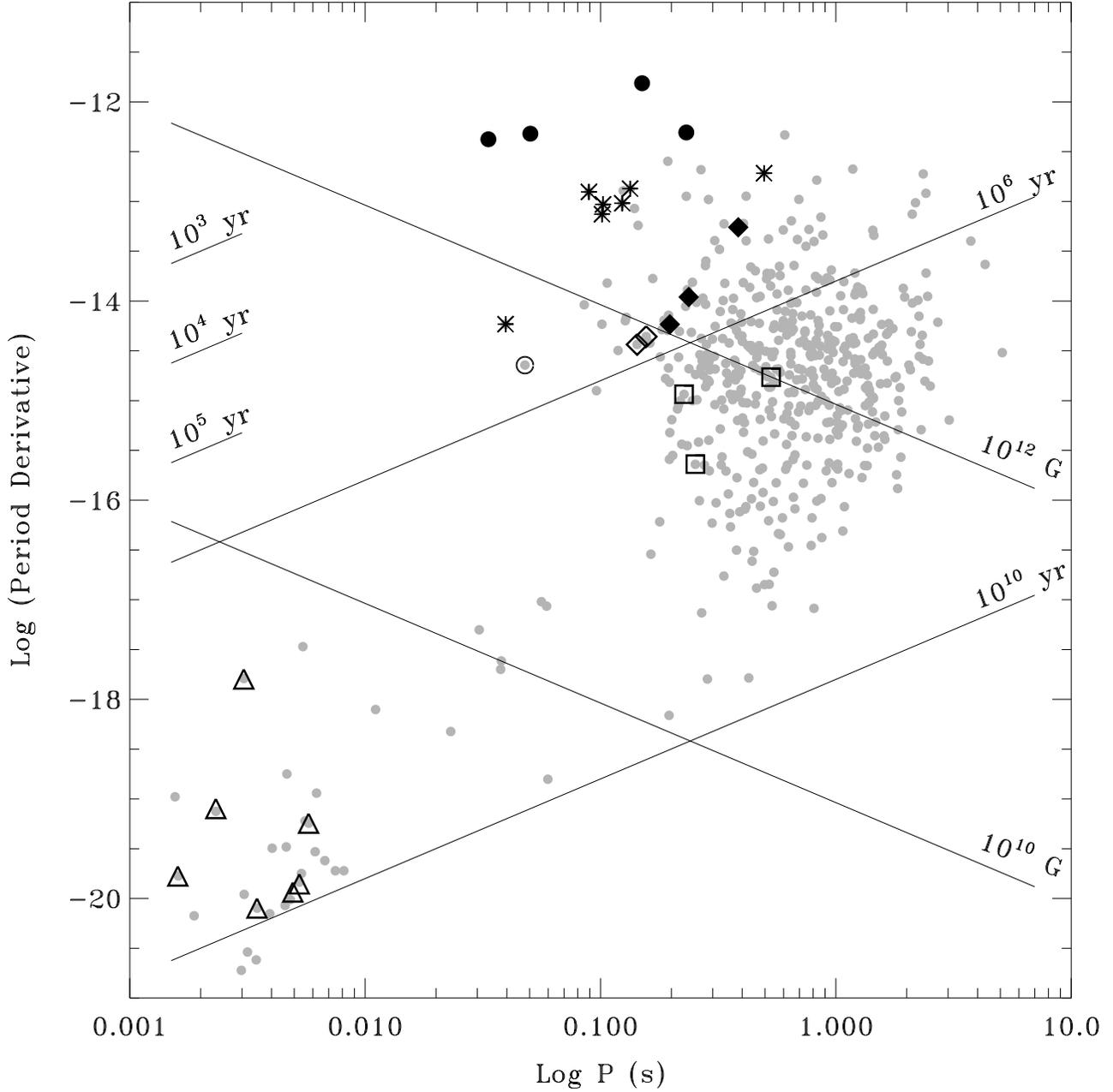,height=17.8cm,clip=}}  
  \caption[]{The sample of rotation-powered pulsars plotted with
  respect to their spin parameters $P, \dot{P}$. Separate from the
  majority of ordinary field pulsars are the millisecond pulsars in
  the lower left corner. X-ray detected pulsars are indicated by symbols
  defined in Fig.4. Lines representing constant ages $\tau=P/2\dot{P}$
  and magnetic field strength $B_\perp=3.3\times 10^{19}(P\dot P)^{1/2}$
  are indicated.} \label{ppdot}
 \end{figure*}

  For a long time the most luminous of all rotation-powered pulsars, the Crab 
  pulsar, had been the only radio pulsar detected at X-ray energies. With  
  increasing sensitivity of the optical, X- and gamma-ray observations the 
  number of pulsars detected in these spectral bands has steadily increased. 
  At present 8 optical, 27 X-ray and 6 gamma-ray detections are known 
  (cf.~Tab.\ref{sym_tab} and Fig.\ref{ppdot}).
  In X-rays, the first big step was taken with the Einstein Observatory which 
  led to the discovery of pulsations from the Crab-like pulsars PSR 0540-69 
  and PSR 1509-58,  while fluxes were found for PSR 0656+14, PSR 1055-52, 
  PSR 1929+10 and PSR 0950+08 (cf.~Seward \& Wang 1988). 
  More recently, ROSAT ($0.1-2.4$ keV) and ASCA ($\sim 0.7-10$ keV) have 
  provided 18 and 2 further detections, respectively, including 8 pulsars 
  for which X-ray pulsations were discovered. In addition, the superior 
  sensitivity and spectral resolution of the ROSAT {\bf P}osition 
  {\bf S}ensitive {\bf P}roportional {\bf C}ounter (PSPC) provided more 
  detailed spectral information for a number of pulsars especially in the 
  soft band ($0.1-0.6$ keV). Using the good angular resolution of the ROSAT 
  {\bf H}igh {\bf R}esolution {\bf I}mager nebular components could be 
  identified and subtracted.  

 \begin{table*}
 \begin{center} 
 \begin{tabular}{c c c l c c}\hline\hline\\[-1ex]
      Pulsar     &       $N_H$           &    observed in   &        \quad count rate in         &   spectral    & Ref.  \\
        {}       &   $[10^{21} cm^{-2}]$   &    range [KeV]   &        total band [cts/s]          &     info      &  {} \\[2ex]\hline\\[-1.5ex]
 $\;$ B$0531+21$ &    $\sim 3$           & $0.5 - 2.4$ & H:$ 17.8   \pm 0.05$               &     pwl       &  tw    \\
 $\;$ B$0833-45$ &  $0.4\pm 0.1$         & $0.1 - 2.4$ & P:$\;\;3.4 \pm 0.02$               &   pwl or bb   &  tw,1  \\
 $\;$ B$0633+17$ &  $0.1\pm 0.04$        & $0.1 - 2.4$ & P:$\;\;5.4 \pm 0.3 \times 10^{-1}$ &    bb+pwl     &  tw,2  \\
 $\;$ B$1706-44$ &  $\sim 5.4$           & $0.5 - 2.4$ & P:$\;\;2.2 \pm 0.2 \times 10^{-2}$ &     pwl       &   3    \\
 $\;$ B$1509-58$ &  $\sim 8  $           & $0.5 - 2.4$ & H:$\;\;3.6 \pm 0.1\times 10^{-3}$  &     pwl       &  tw    \\
 $\;$ B$1951+32$ &  $3.4\pm 1.5$         & $0.5 - 2.4$ & P:$\;\;6.8 \pm 0.3 \times 10^{-2}$ &     pwl       &  tw,4  \\
 $\;$ B$1046-58$ &  $\sim 4$             & $0.5 - 2.4$ & H:$\;\;2.3 \pm 0.4 \times 10^{-3}$ &     no        &  tw    \\
 $\;\;\;$ B$1259-63^*$ &  $3.6\pm 0.5$   & $0.5 - 2.4$ & P:$\;\;3.0 \pm 0.1 \times 10^{-2}$ &     pwl       &  tw    \\
 $\;$ B$1823-13$ &  $\sim 8.2$           & $0.5 - 2.4$ & P:$\;\;8.6 \pm 0.1 \times 10^{-3}$ &     no        &  tw,5  \\
 $\;$ B$1800-21$ &  $\sim 14$            & $0.5 - 2.4$ & P:$\;\;1.5 \pm 0.5 \times 10^{-3}$ &     no        &  tw,6  \\
 $\;$ B$1929+10$ &  $0.8\pm 0.3$         & $0.5 - 2.4$ & P:$\;\;1.17\pm 0.07\times 10^{-2}$ &   pwl or bb   &  tw,7  \\
 $\;$ J$0437-47$ &  $0.08\pm 0.02$       & $0.1 - 2.4$ & P:$\;\;2.04\pm 0.06\times 10^{-1}$ &     pwl       &  tw,8  \\
 $\;$ B$1821-24$ &  $\sim 3$             & $0.5 - 2.4$ & P:$\;\;1.0 \pm 0.2 \times 10^{-2}$ &   ASCA,pwl    &  tw,9  \\
 $\;$ B$0656+14$ &  $0.06\pm 0.01$       & $0.1 - 2.4$ & P:$\;\;1.92\pm 0.03$               & bb+pwl(fixed) &  tw    \\
 $\;$ B$0540-69$ &  $4 (+0.6;-0.4)$      & $0.5 - 2.4$ & H:$\;\;3.7 \pm 0.1\times 10^{-2}$  &     pwl       &  tw,10 \\
 $\;$ J$2124-33$ &  $0.2- 0.5$           & $0.1 - 2.4$ & H:$\;\;2.6 \pm 0.2 \times 10^{-3}$ &     no        &  tw,8  \\
 $\;$ B$1957+20$ &  $\sim 4.5$           & $0.5 - 2.4$ & P:$\;\;4.2 \pm 0.6 \times 10^{-3}$ &     no        &  tw    \\
 $\;$ B$0950+08$ &  $\sim 0.17$          & $0.1 - 2.4$ & P:$\;\;4.9 \pm 0.9 \times 10^{-3}$ &     no        &  tw,11 \\
 $\;$ B$1610-50$ &        {}             &   {}        &             {}                     &    ASCA,no    &    12  \\
 $\;$ J$0538+28$ &  $\sim 1.2$           & $0.5 - 2.4$ & P:$\;\;6.0 \pm 0.1 \times 10^{-2}$ &     no        &  tw    \\
 $\;$ J$1012+53$ &  $\sim 0.07$          & $0.1 - 2.4$ & P:$\;\;5.5 \pm 2.0 \times 10^{-3}$ &     no        &  tw,13 \\
 $\;$ B$1055-52$ &  $0.33\pm 0.1$        & $0.1 - 2.4$ & P:$\;\;3.51\pm 0.05\times 10^{-1}$ & bb+pwl(fixed) &  tw    \\
 $\;$ B$0355+54$ &  $\sim 1.8$           & $0.5 - 2.4$ & P:$\;\;4.0 \pm 1.4 \times 10^{-3}$ &     no        &  tw    \\
 $\;$ B$2334+61$ &  $\sim 3$             & $0.5 - 2.4$ & P:$\;\;1.8 \pm 0.5 \times 10^{-3}$ &     no        &  tw,14 \\
 $\;$ J$0218+42$ &  $\sim 0.5$           & $0.1 - 2.4$ & H:$\;\;2.1 \pm 0.4 \times 10^{-3}$ &     no        &  tw,15 \\
 $\;$ B$0823+26$ &  $\sim 0.4$           & $0.1 - 2.4$ & P:$\;\;1.6 \pm 0.4 \times 10^{-3}$ &     no        &  tw    \\
 $\;$ J$0751+18$ &  $\sim 0.44$          & $0.1 - 2.4$ & P:$\;\;3.6 \pm 0.6 \times 10^{-3}$ &     no        &  tw,16 \\[-1ex]
 \multicolumn{6}{c}{\rule[0mm]{0mm}{0mm}}\\\hline\hline\\[-3ex]
 \end{tabular}
 \caption[]{List of observational properties of the X-ray detected
  rotation-powered pulsars, ordered according to $\dot{E}/4 \pi d^2$.
  The individual columns are as follows:
  1.~Pulsar name,
  2.~column density in units of $10^{21}\;\mbox{cm}^{-2}$,
  3.~energy range in which the soft X-rays have been detected with ROSAT,
  4.~background and vignetting corrected count rate after cleaning the 
     data for solar scattered X-rays and the particle background
     contribution. P,H indicates PSPC and HRI, respectively,
  5.~spectral information; pwl: power-law, bb: black-body spectrum. For PSR 0656+14
     and PSR 1055-52 the photon-index of the power-law spectrum has to be fixed to
     yield good spectral fits with the compound black-body and power-law model.
     We fixed it to $\alpha=-2$. 
     ${}^*$PSR 1259$-$63 was observed at an orbital phase angle of $\sim 13^\circ$ post-apastron.
   Ref: ${}^{tw}$this work, 
        ${}^1$\"Ogelman (1995),
        ${}^2$Halpern \& Ruderman (1993),
        ${}^3$Becker, Brazier \& Tr\"umper (1995),
        ${}^4$Safi-Harb, \"Ogelman \& Finley (1995), 
        ${}^5$Finley \&  Srinivasan (1996), 
        ${}^6$Finley \& \"Ogelman (1994),
        ${}^7$Yancopoulos, Hamilton \& Helfand (1994),
        ${}^8$Becker et al.~(1997b), 
        ${}^9$Danner, Kulkarni \& Thorsett (1994), 
        ${}^{10}$Finley et al.~(1993), 
        ${}^{11}$Manning \& Willmore (1993),
        ${}^{12}$Kawai \& Tamura (1997), 
        ${}^{13}$Halpern (1996), 
        ${}^{14}$Becker, Brazier \& Tr\"umper (1996), 
        ${}^{15}$Verbunt et al.~(1996), 
        ${}^{16}$Becker et al.~(1996). \\[-4.2ex]
    \label{obs_properties}}  
 \end{center}
 \end{table*}

\section{Data analysis}

  The observational data have been reviewed recently by \"Ogelman (1995),
  Tr\"umper (1995) and Becker (1995a,b). However, most of the spectra and
  luminosities were collected from the available literature, resulting in
  a rather inhomogeneous data set. Actually, the quoted soft X-ray luminosity
  depends strongly on the detector type (the PSPC has limited spectral resolution,
  the HRI none), spectral model, distance model, absorption column and photon
  statistics.
  In this paper we present a homogeneous data set obtained by re-analyzing
  most of the ROSAT observations of pulsars in a standard way by making
  power-law and black-body fits to the count rate spectra using EXSAS
  (Zimmermann et al.~1994).
  The absorption columns $N_H$ are deduced from spectral fits, or, if no
  detailed analysis was possible, from HI surveys (Dickey \& Lockman 1990,
  Stark et al.~1992), from the optical extinction (cf.~Becker et al.~1997b
  and references therein) or from the radio dispersion measure $DM$ by
  assuming a mean electron density of $\bar{n}_e=0.03\; \mbox{cm}^{-3}$.
  For the pulsar distances we used the model of Taylor \& Cordes (1993)
  which {\em on average} yields distances correct up to a nominal error
  of about 25\%.
  We used the proper motion corrected period derivatives to compute $\dot{E}$,
  $B_\perp$ and the spin-down age $\tau$ for all those pulsars for which this
  effect is of significance and has been measured (Camilo, Thorsett \& Kulkarni
  1994, Bell et al.~1995).  

\begin{figure} \label{photon_indices_fig}
 \centerline{\psfig{figure=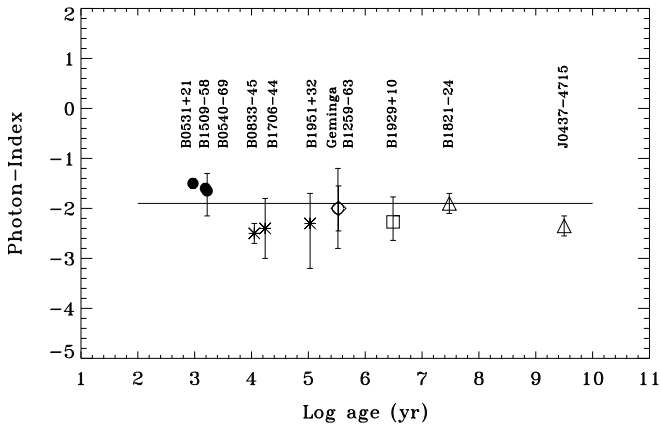,width=9.5cm}}
  \caption[]{The distribution of pulse-phase averaged photon indices  
   vs.~spin-down age for those pulsars which provides spectral 
   information in the soft X-ray domain. The weighted mean of the 
   photon indices is $\bar{\alpha}=-1.9 \pm 0.8$.}
 \end{figure}

   While our data analysis is done in a homogeneous way, the observational data 
  differ significantly from source to source with respect to the total number
  of counts available and the energy range covered. Observational properties
  are summarized in Table \ref{obs_properties}. Sources for which a spectral 
  analysis is meaningless for statistical reasons or which have been observed 
  with the HRI only are labeled in the column {\em spectral info} with {\em no}.
  It turns out that with the exception of the younger pulsars (Crab, PSR 1509-58
  and Vela) all measured pulse-phase averaged photon indices are consistent with 
  $\alpha \approx -2$ (cf.~Figure \ref{photon_indices_fig}). Therefore in calculating 
  luminosities of the power-law components we used the individual photon indices 
  given in Table \ref{photon_indices_tab} and a {\em canonical} value of $\alpha 
  =-2$ where no spectral information was available.\\[-6ex]

\vspace{-5ex}

 \begin{table}
 \begin{center}
 \flushleft \label{photon_indices_tab}
 \caption[]{List of pulse-phase averaged photon-indices}
 \begin{tabular}{c c c}\hline\hline\\[-2ex]
     Pulsar                 &       $\alpha$          &         Ref. $\;$   \\[+1ex]\hline\\[-2ex]
 $\;$ B$0531+21$            &  $-1.5\pm 0.1$          &  Toor \& Seward (1977) \\
 $\;$ B$1509-58$            &  $-1.6\pm 0.1$          &  Kawai et al.~(1992)   \\
 $\;$ B$0540-69$            &  $-1.65^{+0.35}_{-0.5}$ &  Finley et al.(1993)   \\
 $\;\;$ B$0833-45^\ast$     &  $-2.5\pm 0.2$          &      this work         \\
 $\;$ B$1706-44$            &  $-2.4\pm 0.6$          &  Becker  et al.~(1995) \\
 $\;\;$ B$1951+32^\ast$     &  $-2.3^{+0.6}_{-0.9}$   &      this work         \\
 $\;\;\;\;$ B$1259-63^{\ast\ast}$ &  $-2.0\pm 0.8$    &      this work         \\
 $\;$ B$0633+17$            &  $-2.0\pm 0.45$         &      this work         \\
 $\;$ B$1929+10$            &  $-2.27^{+0.5}_{-0.37}$ &      this work         \\
 $\;$ B$1821-24$            &  $-1.9\pm 0.2$          &    Saito et al.(1997)  \\ 
 $\;$ J$0437-47$            &  $-2.35 \pm 0.2$        &    Becker  et al.~(1997b)     \\[+1ex]\hline\hline
\end{tabular}
\end{center}
\par
 \vspace{1ex}
 \renewcommand{\baselinestretch}{1} \noindent \normalsize \small\footnotesize
  ${}^\ast$Photons from within a 35" circle centered on the pulsar.\\
  ${}^{\ast\ast}$ Observed at $13^\circ$ post-apastron.
\end{table}

  The energy fluxes $f_x$ within the 0.1-2.4 keV band have been converted to 
  isotropic luminosities using $L_x=4\pi d^2 f_x$. The X-ray luminosities for the 
  total pulsar emission ($L_{x}^{tot}$), a pulsed component if detected 
  ($L_{x}^{puls}$) and for the pulsar plus nebula ($L_{x}^{pn}$) if a nebula has
  been identified, are given in Table \ref{sym_tab}. 
  We further give there the bolometric luminosities ($L^{\infty}_{bol}$) for the 
  black-body component of Geminga, PSR 0656+14 and 1055-52 (which in the following 
  we will call {\em the three musketeers}) and upper limits for the other pulsars
  assuming a neutron star with a medium stiff equation of state (FP-Model, 
  $M=1.4\;\mbox{M}_\odot$ and $R=10.85$ km).
  In Table \ref{sym_tab} the sources are ordered according to $\dot{E}/4\;\pi\;d^2$ 
  down to  $\sim 1.5 \times 10^{-11}\;\mbox{erg s}^{-1}\mbox{cm}^{-2}$. The Princeton 
  pulsar archive (Taylor et al.~1995) contains 71 pulsars above this limit, including 
  45 objects which have not been detected in X-rays yet\footnote{We note that one 
  X-ray detected pulsar was not listed in the available version of the Princeton 
  pulsar catalog.}. 12 of them have been observed by ROSAT in pointed observations
  but were not detected while 10 were in the ROSAT field of view serendipitously. 
  We have derived $1\sigma$ upper limits for 19 of the 22 pulsars
  applying the procedure described above. The corresponding upper limits for their 
  luminosity are given in Tab.\ref{upper_lim_tab} and in Fig.\ref{Lx_Edot}, which 
  we shall discuss later. 

 \begin{table*}
 \centerline{\psfig{figure=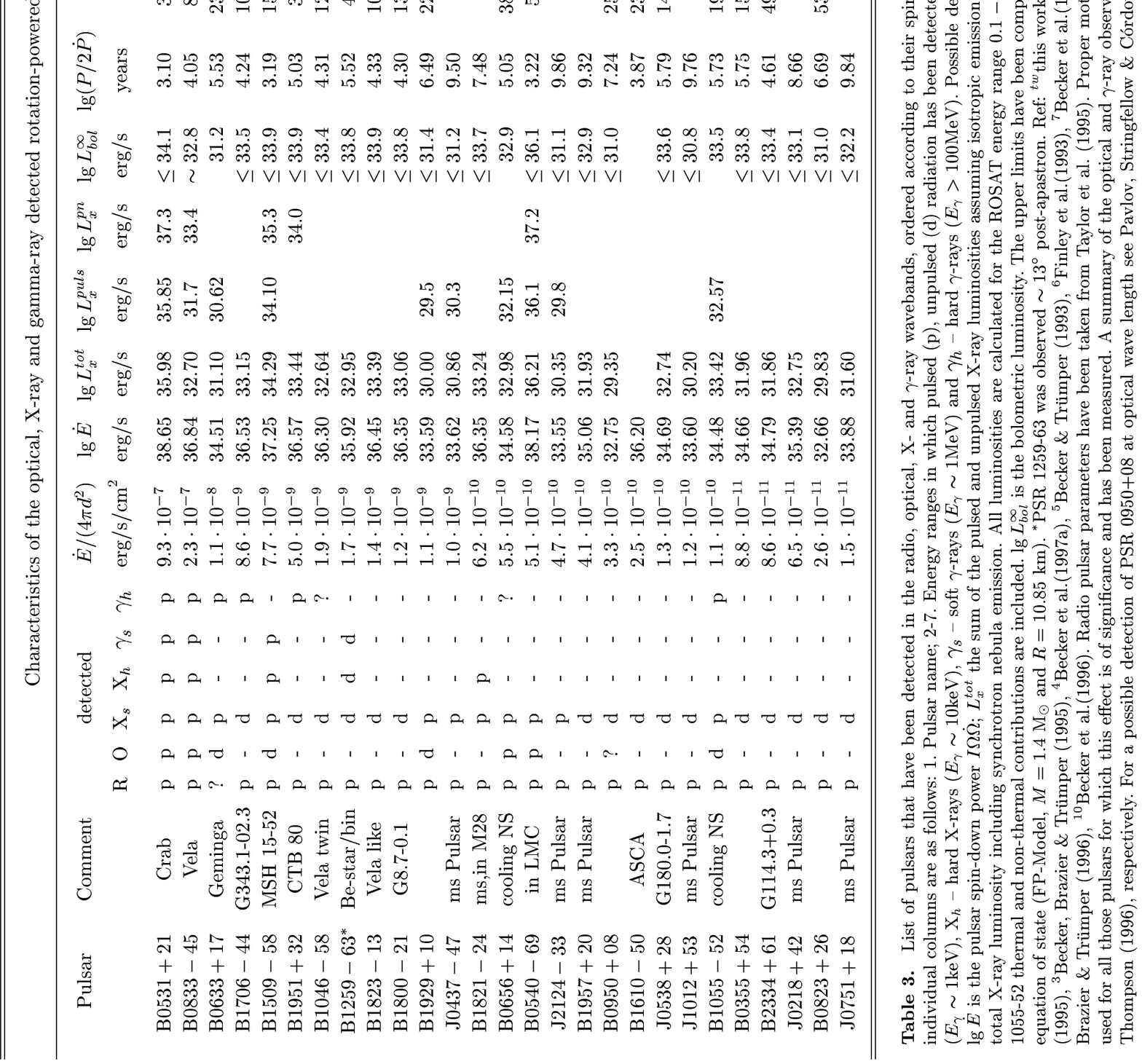,angle=90,width=17.5cm,height=23.5cm}}
 \refstepcounter{table} \label{sym_tab}
 \end{table*}

\begin{table*}
\begin{center}
\begin{tabular}{c l c c c c c r c c c}\hline\hline\\[-1ex]
   Pulsar   & $\dot{E}/(4\pi d^2)$ & \Edot & $\lg(P/2\dot{P})$  &  $P$   &   \Pdot     &  D   & $\lg B_\perp$ &      $N_H$           &    count rate     & $L_x$   \\
    {}      & $\mbox{erg/s/cm}^2$  & erg/s &      years         &   ms   & s s$^{-1}$  & kpc  &   Gauss       &   $10^{21} cm^{-2}$  &      cts/s        &  erg/s  
\\\\[-1ex]\hline\\[-1.2ex]
 B$1757-24$  &$1.1\cdot 10^{-9}$  & 36.41 &  4.19 & 124.87 &   127.89    & 4.61 &   12.61       &     $\sim 9.2$       &  H: $\le 0.0014$ & $\le 33.17$   \\
 B$1937+21$  &$7.1\cdot 10^{-10}$ & 36.04 &  8.37 &  1.557 &     1.1 E-4 & 3.60 &    8.61       &     0.28 - 2.2       &  H: $\le 0.0007$ & $\le 32.10$   \\
 B$1727-33$  &$5.8\cdot 10^{-10}$ & 36.09 &  4.42 & 139.45 &    85.00    & 4.24 &   12.54       &     $\sim 7.9$       &  P: $\le 0.0020$ & $\le 32.73$   \\
 B$1257+12$  &$4.1\cdot 10^{-10}$ & 33.73 &  9.50 &   6.22 &     3.3E-5  & 0.62 &    8.66       &     $\sim 3.2$       &  P: $\le 0.0011$ & $\le 29.99$   \\
 B$0114+58$  &$4.0\cdot 10^{-10}$ & 35.34 &  5.44 & 101.44 &     5.84    & 2.14 &   11.89       &     $\sim 1.5$       &  P: $\le 0.0033$ & $\le 31.84$   \\
 B$0740-28$  &$3.4\cdot 10^{-10}$ & 35.16 &  5.20 & 166.76 &    16.811   & 7.80 &   12.23       &     $\sim 2.0$       &  P: $\le 0.0020$ & $\le 32.79$   \\
 B$1853+01$  &$3.3\cdot 10^{-10}$ & 35.63 &  4.31 & 267.40 &   208.40    & 3.30 &   12.88       &     $\sim 3.0$       &  H: $\le 0.0022$ & $\le 32.63$   \\
 B$1830-08$  &$1.5\cdot 10^{-10}$ & 35.77 &  5.17 &  85.28 &     9.17    & 5.67 &   11.95       &     $\sim 12.6$      &  P: $\le 0.0017$ & $\le 33.16$   \\
 $\;\;\;$B$1820-30$A & $1.1\cdot 10^{-10}$& 35.92 &  7.41 &   5.44 &     3.3E-3  & 8.00 &    9.64       &     $\sim 2.7$       &         {}       &     {}        \\
 J$1908+07$  &$8.4\cdot 10^{-11}$ & 33.53 &  6.61 & 212.35 &     8.25E-1 & 0.58 &   11.63       &     $\sim 0.34$      &  P: $\le 0.0045$ & $\le 30.57$   \\
 B$1737-30$  &$6.4\cdot 10^{-11}$ & 34.91 &  4.32 & 606.69 &   465.30    & 3.28 &   13.23       &     4.7 - 9.5        &  P: $\le 0.0023$ & $\le 32.36$   \\
 B$1620-26$  &$5.9\cdot 10^{-11}$ & 34.36 &  8.35 &  11.07 &    7.90E-4  & 1.80 &    9.48       &     $\sim 1.9$       &  P: $\le 0.0010$ & $\le 31.66$   \\
 J$2322+20$  &$4.8\cdot 10^{-11}$ & 33.15 & 10.32 &   4.80 &     4E-6    & 0.78 &    8.11       &     $\sim 0.4$       &  P: $\le 0.0006$ & $\le 29.99$   \\
 J$2019+24$  &$4.6\cdot 10^{-11}$ & 33.11 & 10.43 &   3.93 &     2E-6    & 0.91 &    8.00       &     $\sim 0.5$       &  P: $\le 0.0007$ & $\le 30.25$   \\
 B$1822-09$  &$3.7\cdot 10^{-11}$ & 33.66 &  5.37 & 768.98 &    52.36    & 1.01 &   12.81       &     $\sim 0.6$       &  P: $\le 0.0014$ & $\le 30.65$   \\
 J$0631+10$  &$3.4\cdot 10^{-11}$ & 35.24 &  4.64 & 287.75 &   104.62    & 6.56 &   12.74       &     $\sim 3.9$       &  P: $\le 0.0051$ & $\le 33.23$   \\
 B$1534+12$  &$3.2\cdot 10^{-11}$ & 33.24 &  8.30 &  37.90 &     2.4E-3  & 0.68 &    9.98       &     $\sim 0.4$       &  P: $\le 0.0019$ & $\le 30.35$   \\
 B$1356-60$  &$2.9\cdot 10^{-11}$ & 35.08 &  5.50 & 127.50 &     6.33    & 5.91 &   11.96       &     $\sim 9.1$       &        {}        &     {}        \\
 B$1754-24$  &$2.7\cdot 10^{-11}$ & 34.60 &  5.46 & 234.09 &    13.00    & 3.50 &   12.25       &     $\sim 5.5$       &        {}        &     {}        \\
 B$0611+22$  &$2.4\cdot 10^{-11}$ & 34.80 &  4.95 & 334.92 &    59.63    & 4.72 &   12.65       &     $\sim 3.0$       &  P: $\le 0.0011$ & $\le 32.20$   \\
 B$1828-10$  &$2.3\cdot 10^{-11}$ & 34.55 &  5.03 & 405.03 &    60.04    & 3.63 &   12.70       &     $\sim 5.0$       &  P: $\le 0.0022$ & $\le 32.24$   \\
 B$1742-30$  &$1.6\cdot 10^{-11}$ & 33.93 &  5.74 & 367.43 &    10.66    & 2.08 &   12.30       &     $\sim 2.7$       &  P: $\le 0.0007$ & $\le 31.18$   \\
 \multicolumn{11}{c}{\rule[0mm]{0mm}{0mm}}\\[-1ex]\hline\hline\\[-3ex]
 \end{tabular}

\caption[]{List of undetected pulsars observed in pointed and serendipitous ROSAT observations, ordered according to $\dot{E}/4 \pi d^2$.
 The individual columns are as follows:
 1.~Pulsar name,
 2-10.~the pulsar's spin-down flux density at earth, its braking energy, characteristic age, period and period 
 derivative, distance and magnetic dipole component, column density from dispersion measure and count rate upper 
 limits. P,H indicates PSPC and HRI, respectively, 
 11.~X-ray luminosity within 0.1-2.4 keV, assuming isotropic emission.
 We note that for PSR 1820-30A, 1356-60 and 1754-24 the data quality wasn't sufficient to deduce a useful upper 
 limit (e.g.~source located at the edge of the detector's field of view with an exposure time of less than a 
 kilosecond).
 \label{upper_lim_tab}}
 \end{center}
 \end{table*}

\section{The X-ray emission properties of pulsars}

  Before taking a synoptic look at the sample of the 26 pulsars detected in the
  ROSAT band a few additional remarks on their X-ray emission characteristics are
  appropriate. To this end we group the whole sample into five classes:

\subsection{The Crab-like pulsars}
  It is well established that in the young Crab-like pulsars with ages 
  $\le 2000$ years magnetospheric emission dominates. In the case of 
  the Crab pulsar, for example, at least $\sim 75\%$ of the total soft 
  X-ray flux is magnetospheric emission (Becker \& Aschenbach 1995) 
  characterized by a power-law spectrum and sharp pulses. With the ROSAT 
  HRI an upper limit for the unpulsed flux within the 0.1-2.4 keV range 
  can be deduced from the DC level of the soft X-ray pulse profile. The 
  latter can be taken as an upper limit for the thermal flux from the Crab 
  pulsar which is marginally consistent with standard cooling models
  (Becker \& Aschenbach 1995, cf.~Fig.\ref{cooling}).
  The same holds for PSR 1509-58 which from our ROSAT HRI observation
  is found to have a pulsed fraction of $65\pm 4$\% (Becker et al.~1997a)
  and a soft X-ray pulse which is aligned with the hard pulses detected 
  at $20-170$ keV by BATSE and OSSE (Ulmer et al.~1994). The pulsar's 
  count rate deduced from the DC-level of the soft X-ray light curve can 
  be taken as an upper limit for the thermal flux (cf.~Tab.3 and 
  Fig.\ref{cooling}).

\subsection{The Vela-type pulsars}
  Different from the Crab-like pulsars are pulsars in the age bracket 
  $\sim 10^4-10^5$ years (e.g.~the Vela-pulsar, PSR 1706-44, 1046-58, 
  1800-21, 1823-13 and 1951+32). They exhibit strong steady emission 
  from a  pulsar-powered synchrotron nebula combined with a small pulsed 
  contribution of magnetospheric or thermal origin dominating the 
  emission in the range $\sim 0.1-0.5$ keV (\"Ogelman, Finley \& 
  Zimmerman 1993; Becker \& Tr\"umper 1996). 
  X-ray pulses are only detected for the Vela pulsar. The other Vela-type 
  pulsars are more distant and suffer from photoelectric absorption which
  prevents the detection of their soft pulses in the presence of the 
  dominant nebula emission (Becker \& Tr\"umper 1996).
  Apart from the Vela pulsar itself there is no spectral information for 
  this sources below 0.5 keV (cf.~Table \ref{obs_properties}) which means 
  that their black-body component is invisible because of photoelectric 
  absorption.

\subsection{The cooling neutron stars}

  Our analysis confirms the results obtained earlier by several authors
  (e.g.~\"Ogelman \& Finley 1993; Becker et al.~1993; Mereghetti \& 
  Colpi 1996; Halpern \& Wang 1997) on the three middle age pulsars Geminga,
  PSR 0656+14 and 1055-52. The X-ray emission properties of these
  pulsars are characterized by a dichotomy, i.e.~the spectra are best
  described by a two-component model in which the soft emission is
  represented by a black-body spectrum and the hard component either
  by a thermal spectrum or by a power-law. The existence of two spectral
  components is also confirmed by phase-resolved analysis. All three
  pulsars show a phase shift of $\sim 100^\circ$ and a change in the 
  pulsed fraction from $\sim 10-30\%$ below a transition point of 
  $0.5-0.6$ keV, rising up to $\sim 20-65\%$ above. The X-ray pulse
  profile for both the soft and the hard components is found to be
  sinusoidal. 

  The soft thermal emission is assigned to be cooling emission 
  from the neutron star's surface. The modulation of this emission 
  can be explained by non-uniformities in the surface temperature 
  due to the presence of a strong magnetic field which gives rise 
  to an anisotropic heat flow in the neutron star's outer layers.
  The radius of the emitting area obtained by using $R_{bb}=d/T^2\,
  \sqrt{f_{bol}/\sigma}$, in  which $f_{bol}$ and $T$ denotes the fitted 
  bolometric flux and temperature of the pulsar's soft component and 
  $\sigma$ the Stefan-Boltzmann constant, are found to be close to 
  the canonical neutron star radius of 10 km.

  The hard spectral components of these pulsars can be interpreted 
  as magnetospheric emission or thermal radiation from polar hot spots.
  Because of bandwidth limitations a distinction between these possibilities
  cannot be made on the basis of ROSAT PSPC spectra.
  But ASCA observations have shown recently that the hard component of Geminga
  is characterized by a power-law implying a magnetospheric origin (Halpern \&
  Wang 1997). Similar results have been reported for PSR 0656+14 and 1055-52
  (Greiveldinger et al.~1996), although the limited photon statistics in these
  cases prevent a clear distinction between a thermal and non-thermal origin 
  of the emission.

\begin{figure}
\centerline{\psfig{figure=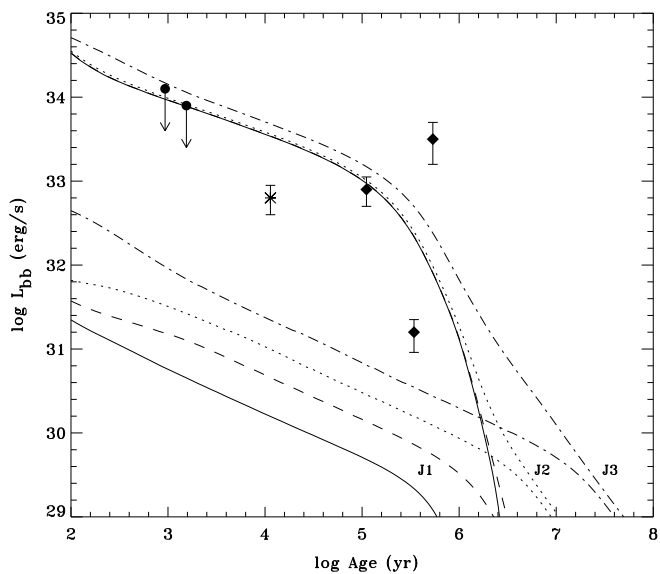,width=9.4cm,clip=}}
\caption[]{Bolometric luminosities $L^\infty_{bol}=4\,\pi\,d^2\,f_{bol}$ for
  the three musketeers (PSR 0656+14, Geminga and 1055-52) and the Vela pulsar
  as a function of the pulsar's characteristic age. Also shown are the upper
  limits for the Crab pulsar and PSR 1509-58 (cf.~Table \ref{sym_tab}). Symbols
  are defined in Fig.\ref{Lx_Edot}. For comparison, the cooling curves for
  standard and accelerated neutron star cooling (FP-Model) including frictional
  heating with strong (J3), weak (J2) and super-weak (J1) pinning of crustal
  superfluid vortex lines are shown (Umeda et al.~1993). We note that the given
  error bars indicate $1\sigma$ errors as obtained from spectral fits. Distance
  uncertainties which might contribute a much larger error are not taken into
  account.}\label{cooling}
\end{figure}

  In Figure \ref{cooling} we have plotted the bolometric black-body luminosities
  $L^{\infty}_{bol}=4\,\pi\,d^2\,f_{bol}$ for the three musketeers and Vela
  as obtained from spectral fits.
  The figure demonstrates that the thermal emission from PSR 0656+14 and PSR
  1055-52 is consistent with the prediction of standard cooling for a neutron 
  stars with a medium stiff equation of state, while Vela and Geminga are 
  located below the standard cooling curves.

  We note that for the three musketeers optical (Pavlov, Stringfellow \& Cordova 
  1996; Mignani, Caraveo \& Bignami 1997a; Shearer et al.~1996a) 
  and for Geminga and PSR 0656+14\/ EUV emissions have been found (PSR 1055-52 has 
  not yet been observed in the EUV regime). The ROSAT and EUVE spectra can be 
  fitted by a single modified black-body component using photospheric opacities 
  (Zavlin et al.~1995).
  In the case of Geminga the optical emission lies close to the Rayleigh-Jeans 
  extrapolation of the X-ray/EUV spectrum  while for PSR 0656+14 and PSR 1055-52
  the optical flux is higher than that by about one order of magnitude (Shearer 
  et al.~1996b; Mignani, Caraveo \& Bignami 1997b,  Kurt et al.~1997). 
  In Geminga an optical emission feature was reported at $\sim 6000$ \AA$\;$ 
  (Bignami et al.~1996) which has been attributed to proton cyclotron emission  
  from a photospheric plasma. 

  Besides PSR 0656+14, Geminga and PSR 1055-52 two more pulsars are classified as
  cooling neutron stars: PSR 0538+28 and 0355+54. Both have spin parameters
  similar to the one observed for Geminga and 1055-52 and both appear to be good
  candidates for gamma-ray pulsars. However, the sources are approximately a
  factor of 10 more distant than Geminga and the limited photon statistics did
  not allow a spectral or temporal analysis so that their classification is not
  yet based on X-ray observations.

  The three cooling neutron star candidates located in the supernova remnants 
  RCW 103, PKS 1209-51/52 and Puppis A (cf.~Caraveo, Bignami \& Tr\"umper 1996
  and references therein) will not be considered here.

\subsection{Pulsars old and close in space}

 Besides the young pulsars and those whose surface cooling is visible 
 in the X-ray wave band, ROSAT has detected X-ray emission from further 
 three pulsars: PSR 1929+10, 0950+08 and 0823+26. All are characterized
 by a spin-down age of $\sim 0.2-3 \times 10^7$ years and a close 
 distance of $\sim 0.12 - 0.38$ kpc. Pulsed X-ray emission, however,
 could only be detected from PSR 1929+10 (Yancopoulos, Hamilton \& Helfand 
 1994). Its X-ray pulse profile is very broad with a single pulse stretching  
 across almost the entire phase cycle. Spectral information is only available
 for 1929+10. We found that both a power-law and a black-body spectrum fits 
 the data equally well. The results of our spectral fits are given in Table 
 \ref{obs_properties}$-$\ref{sym_tab}.

\begin{figure*}
\centerline{\psfig{figure=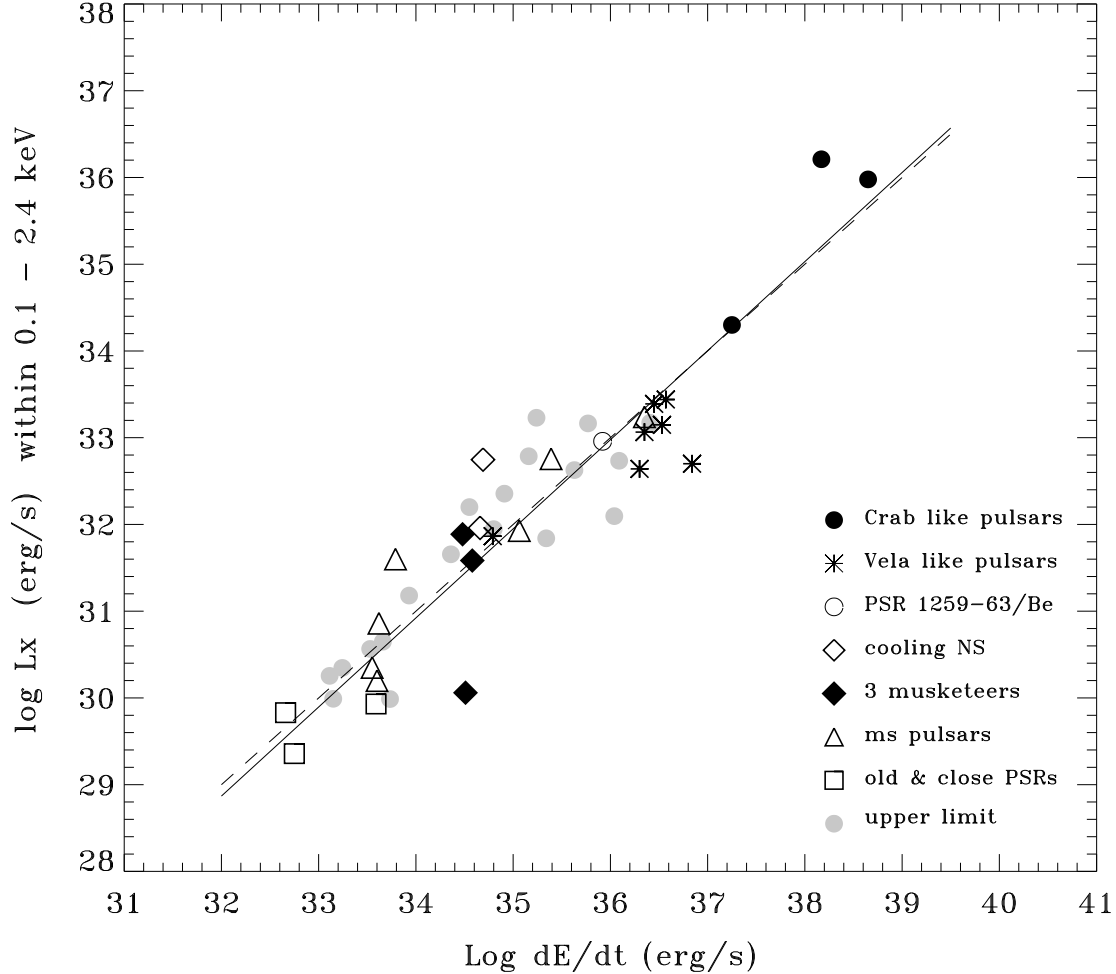,width=16cm,clip=}} 
\caption[]{X-ray luminosity vs.~spin-down energy for all rotation-powered
 pulsar detected by ROSAT as of October 1996. We note that for the three
 musketeers Geminga, PSR 0656+14 and 1055-52 the low energy thermal
 component has been subtracted from the data. The solid line represents
 $L_x(\dot{E}) \propto \dot{E}^{1.03}$, the dashed line $L_x(\dot{E})
 = 10^{-3}\times \dot{E}$. $1\sigma$ upper limits for the X-ray
 luminosity of all pulsars which have a $\dot{E}/4 \pi d^2 \ge
 1.5\times 10^{-11}\;\mbox{erg s}^{-1} \mbox{cm}^{-2}$ and have been
 observed but were not detected in pointed and serendipitous ROSAT
 observations are indicated.}\label{Lx_Edot} 
  \vspace{1.5ex}
\end{figure*}

\subsection{The millisecond pulsars \label{ms_psr}}
  The millisecond pulsars form a separate group among the rotation-powered 
  pulsars. They are distinguished by their small spin periods $(P\le 20$ ms) 
  and high rotational stability ($dP/dt\approx 10^{-18} - 10^{-21}$) and, 
  consequently, they are very old objects with spin-down ages of typically 
  $10^9-10^{10}$ years and magnetic field strengths of the order of 
  $10^8 - 10^{10}$ G.
  Before ROSAT nothing was known about the X-ray emission properties of this 
  class of pulsars. 
  Their location in the $L_x-\tau$ diagram (cf.~Fig.5) had initially led to the
  suggestion that their radiation is a mixture of spin-modulated thermal emission 
  from heated polar caps possibly combined with unpulsed emission from a pulsar 
  wind or a plerion (Becker \& Tr\"umper 1993). The PSPC spectrum of PSR 
  J0437$-$4715 can be fitted by a power-law with a photon-index of $\approx -2.4$. 
  A feature in the energy dependent pulsed fraction of this source which had been 
  interpreted as evidence for an additional weak thermal component (Becker \& 
  Tr\"umper 1993) has disappeared in the present re-analysis (Becker \& Tr\"umper 
  et al.~1997b).
  Recent ASCA observations of the millisecond pulsar PSR 1821-24 in M28 led 
  to the detection of X-ray pulses up to 10 keV (Saito et al.~1997). The 
  spectrum is a power-law with a pulse-phase averaged photon-index of $\approx
  -2$, indicative of magnetospheric emission.
  The recent discovery of pulsed soft X-rays from the isolated millisecond 
  pulsar PSR J2124-3358 (Becker et al.~1997b) which has spin parameters similar 
  to that of PSR J0437-4715 and a pulse profile similar to the one observed in 
  the radio channel at 480 MHz (Bailes et al.~1997) strongly suggests that the 
  X-ray emission of millisecond pulsars is of magnetospheric origin. The next  
  paragraph will further support this interpretation.

\begin{figure*}
\centerline{\psfig{figure=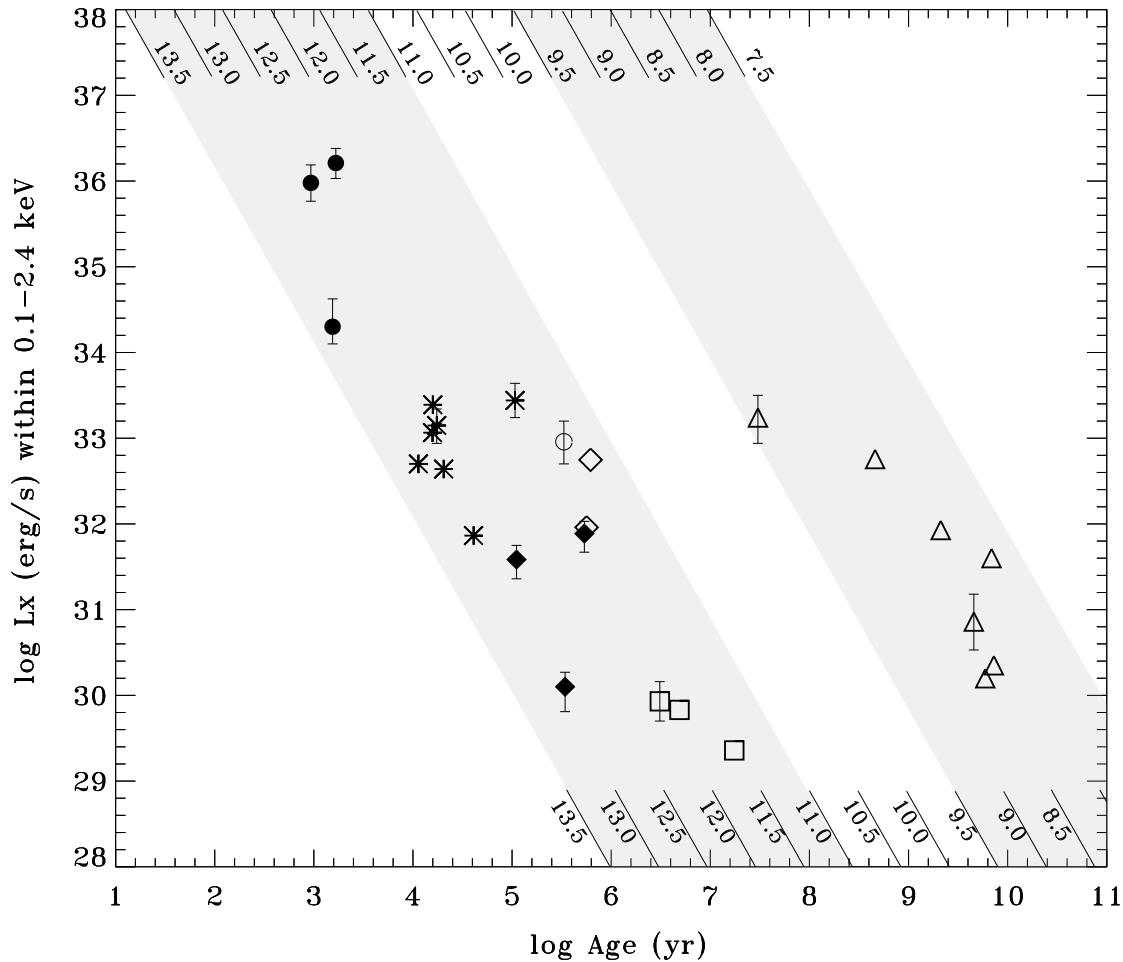,width=16cm,clip=}} 
\caption[]{X-ray luminosities of the ROSAT detected rotation-powered pulsars
 within $0.1-2.4$ keV as a function of the pulsar's characteristic age. The
 plotted luminosities have been derived from power-law spectra and can be
 attributed to magnetospheric emission.  A thermal component as detected for
 the three musketeers and Vela has been subtracted.  Curves corresponding to
 $L_x=\eta_x \;\dot{E}$\/ for $\log B_\perp=7.5, \ldots, 13.5\; \mbox{G}$ are
 indicated. Symbols are defined in Fig.\ref{Lx_Edot}.} \label{LxAge}
\end{figure*}

\section{Synoptic view and discussion}

  Figure \ref{Lx_Edot} shows the X-ray luminosities of all 26 pulsars measured in
  the ROSAT band as a function of their rotational energy loss $\dot{E}$. We note 
  that the young but distant pulsar PSR 1610-50 has not yet been observed by ROSAT 
  and the sparse ASCA data provide only a rough flux estimate (Kawai, Tamura \&
  Shibata 1996) so that we have excluded this source from our analysis.
  In the case of the {\em three musketeers} we have plotted the luminosities 
  derived for the power-law component (i.e.~$\log L_x = 30.06, 31.58, 31.88$ 
  erg/s for Geminga, PSR 0656+14 and 1055-52, respectively).

  Fig.\ref{Lx_Edot} shows a rather close linear correlation between the
  X-ray luminosity and the spin-down power. Fitting a $L_x (0.1-2.4\; 
  \mbox{\footnotesize keV}) = a\;\dot{E}^b$ distribution to the data we 
  find $b=1.03 \pm 0.08$ and a correlation coefficient of $r = 0.946$.
  We note that Seward \& Wang (1988) and \"Ogelman (1995) found for much
  smaller samples $b = 1.39$ and $b = 1.35 \pm 0.25$ respectively, which 
  does not deviate significantly from our value (cf.~also Possenti, 
  Mereghetti \& Stella 1996). Adopting $b=1$ the efficiency of conversion 
  is $\eta_x = L_x/\dot{E}= 10^{-3}$ for the X-rays in the ROSAT band, 
  indicated by the dashed-line in Fig.\ref{Lx_Edot}.

  It is remarkable how narrow the $L_x(\dot{E})$ distribution is in view of 
  the various effects which are expected to widen it: uncertainties in the 
  pulsar distances and the absorption column $N_H$, spread in the orientations 
  of magnetic and rotational axes versus the line of sight and spread in the 
  moments of inertia. These effects are expected to account for a large fraction 
  of the scatter leaving little room for {\em intrinsic fluctuations} of the 
  X-ray efficiency. 
  
  The strong correlation suggests that the prime energy source of the X-ray emission
  is the pulsar's rotational energy. The conversion into X-rays can proceed via 
  different channels: (1) magnetospheric processes or (2) emission from hot 
  spots which both produce pulsed X-rays, (3) pulsar wind or synchrotron nebular 
  emission which lead to unpulsed X-rays. In view of the fact that pulsed emission 
  has been measured for only 11 of the 26 pulsars (cf.~Table \ref{sym_tab}) and 
  unpulsed components may not be spatially resolved because of limited angular 
  resolution (and photon statistics) it is difficult to assess the relative importance
  of the different emission processes.
  
  But in particular the millisecond pulsars showing pulsed emission obey the same linear
  relationship with the same X-ray efficiency as the Crab type pulsars. This indicates that
  their emission is mainly due to magnetospheric processes in line with the occurrence
  of power-law spectra and  the similarity of radio/X-ray pulse profiles as discussed in
  section \ref{ms_psr}. We cannot exclude that some additional components from a polar hot 
  spot contributes (cf.~Zavlin \& Pavlov 1997) but this cannot be the dominating component
  for most pulsars. 
  Only for the three middle aged pulsars PSR 0656+14, Geminga and PSR 1055-52 and
  probably for the Vela-pulsar is an additional thermal component detected which 
  can be attributed to photospheric emission from the neutron stellar surface. 

  Another way to look to the sample of the X-ray detected rotation-powered pulsars is
  shown in Figure \ref{LxAge} where the X-ray luminosity in the ROSAT band is plotted
  as a function of the pulsar's characteristic age.
  Assuming the standard picture of magnetic braking we can write
   \begin{equation}
       L_x = \eta_x \dot{E} \propto R^6\, B_\perp^2 \, \Omega^4 
   \end{equation}     

 \noindent
  where $B_\perp$, $R$, and $\Omega$ are the polar magnetic field strength, radius 
  and angular velocity of the neutron star, respectively. Using $I = 10^{45}\,
  \mbox{g cm}^2$, $R =10^6\,\mbox{cm}$ and making the standard assumption that the 
  present angular velocity $\Omega$ is small compared with the initial one we find
   \begin{equation}
     L_x = \eta_x \dot{E} = const.\, B_\perp^{-2} \, \tau^{-2}
   \end{equation}

 \noindent
  The corresponding $L_x(\tau)$ curves for different magnetic field strength
  are indicated in Fig.\ref{LxAge}. We note that all pulsars lie close to the 
  braking curves calculated for their respective polar magnetic fields. Of 
  course, this message is fully equivalent to that of Fig.\ref{Lx_Edot}, 
  namely that the X-ray luminosity is proportional to the rotational energy loss.

\section{Conclusion}
 Although {\em ordinary field} pulsars and millisecond pulsars form
 well-separated populations in the $L_x(\tau)$ diagram they obey the same
 $L_x \propto \dot{E}$ correlation. The surprisingly close correlation
 between $L_x$ and $\dot{E}$ strongly suggests that the bulk of the observed
 X-rays is emitted at the expense of rotational energy. 
 The fact that the measured X-ray spectra show largely power-laws at $E\ge 
 0.5$ keV indicates that the main emission process is a magnetospheric one, 
 just as in case of the radio and gamma ray emission.

 It is an old question whether the incoherent high energy emission of pulsars
 originates in the co-rotating inner magnetosphere or near the light cylinder.
 In the latter case one expects an intensity dependence on the angular frequency 
 $L_{inc} \propto \Omega^n$ with $n \ge 10$ (Beskin, Gurevich \& Istomin 1993). 
 This is due to the simple fact that the radiation from the light cylinder
 must depend on the total ''particle luminosity'' $L_p \sim B_\perp^2\;\Omega^4$ and
 on the electric and magnetic fields at the light cylinder which scale with 
 $\sim B_\perp \;\Omega^3$. Beskin, Gurevich \& Istomin (1993), for example, find 
 $L_{inc} \sim B_\perp^5\,\Omega^{11}$. 
 In contrast, we obtain $L_x \propto B_\perp^2\;\Omega^4$, assuming the validity 
 of magnetic braking. It thus appears that the bulk of the X-rays observed in the
 sample of 27 pulsars originates in the co-rotating magnetosphere, e.g.~by
 Compton scattering processes (Tr\"umper, Supper \& Becker 1997).

\begin{acknowledgements}
  The ROSAT project is supported by the Bundesministerium f\"ur Bildung, 
  Wissenschaft, Forschung und Technologie (BMBW) and the Max-Planck-Society 
  (MPG). We thank our colleagues from the MPE ROSAT group for their support.
  We acknowledge discussions with G.G.~Pavlov and V.E.~Zavlin as well as
  useful comments of the referee I.~Grenier.
\end{acknowledgements}

\end{document}